\documentclass[aps,prd,groupedaddress,preprint,eqsecnum,nofootinbib]{revtex4}
\usepackage{graphicx,epsf,amssymb,amsbsy,amsfonts,amssymb,amsmath}
\usepackage[usenames, dvipsnames]{color}
\newif\ifdraft
\drafttrue
\newif\ifpreprint
\preprinttrue

\newcommand{\CP}{\mathcal{P}}
\newcommand{\ZZ}{\mathbbm Z}
\def\nn{\nonumber}
\def\tree{{\rm tree}}
\def\pol{\varepsilon}
\def\lr{\leftrightarrow}
\def\P{{\rm P}}
\def\NP{{\rm NP}}
\def\I{{\cal I}}
\def\spa#1.#2{\left\langle#1\,#2\right\rangle}
\def\spb#1.#2{\left[#1\,#2\right]}
\def\Tr{\, {\rm Tr}}

\def\NeqEight{{{\cal N}=8}}
\def\NeqFour{{{\cal N}=4}}
\def\del{\partial}
\def\be{\begin{equation}}
\def\ee{\end{equation}}
\def\bea{\begin{eqnarray}}
\def\eea{\end{eqnarray}}
\def\ba{\begin{eqnarray}}
\def\ea{\end{eqnarray}}
\def\Sect#1{Section~{\ref{#1}}}

\def\eqn#1{eq.~(\ref{#1})}

\def\eqns#1#2{eqs.~(\ref{#1}) and~(\ref{#2})}

\def\Fig#1{Fig.~{\ref{#1}}}

\def\Tab#1{Table~{\ref{#1}}}
\def\tree{{\rm tree}}
\def\oneloop{{\rm 1\hbox{-}loop}}
\def\twoloop{{\rm 2\hbox{-}loop}}
\def\threeloop{{\rm 3\hbox{-}loop}}

\def\e{\epsilon}

\usepackage{bbm}

\begin{document}
\hfuzz 20pt

\ifpreprint
UCLA/11/TEP/104 $\null\hskip0.2cm\null$ \hfill 
SU-ITP-11/08 $\null\hskip0.2cm\null$ \hfill 
SLAC--PUB--14380 $\null\hskip0.2cm\null$ \hfill
CERN--PH--TH/2011-030 $\null\hskip0.2cm\null$ \\
\hbox{~~~~}Saclay--IPhT--T11/030
\fi

\title{Amplitudes and Ultraviolet Behavior of
${\cal N}$\,=\,\,8 Supergravity\footnote{%
Invited talk presented by L.D. at XVI$^{\rm th}$ European Workshop on
String Theory, Madrid, June 14-18, 2010}}

\author{Z.~Bern${}^a$, J.~J.~M.~Carrasco${}^b$, L.~J.~Dixon${}^{c,d}$,
 H.~Johansson${}^e$ and  R.~Roiban${}^f$ }

\affiliation{
${}^a$Department of Physics and Astronomy, UCLA, Los Angeles, CA
90095-1547, USA\\
${}^b$Department of Physics, Stanford University, Stanford, CA 94305, USA\\
${}^c$SLAC National Accelerator Laboratory, Stanford University,
Stanford, CA 94309, USA\\
${}^d$Theory Group, Physics Department, CERN, CH--1211 Geneva 23, 
    Switzerland\\
${}^e$Institut de Physique Th\'eorique, CEA--Saclay,
          F--91191 Gif-sur-Yvette cedex, France\\
${}^f$Department of Physics, Pennsylvania State University,
           University Park, PA 16802, USA
}

\date{March, 2011}

\begin{abstract}
In this contribution we describe computational tools that permit
the evaluation of multi-loop scattering amplitudes in $\NeqEight$
supergravity, in terms of amplitudes in $\NeqFour$ super-Yang-Mills
theory.  We also discuss the remarkable ultraviolet
behavior of $\NeqEight$ supergravity, which follows from these amplitudes,
and is as good as that of $\NeqFour$ super-Yang-Mills theory through at
least four loops.
\end{abstract}

\maketitle


\section{Introduction}
\label{IntroSection}

It is well known that quantum gravity is non-renormalizable by power counting,
due to the dimensionful nature of Newton's constant, $G_N = 1/M_{\rm Pl}^2$.
String theory cures these divergences by introducing a new length scale,
related to the string tension, at which particles are no longer point-like.
The question we wish to address in this contribution is whether a 
non-point-like theory is actually necessary for perturbative finiteness.
Perhaps with enough symmetry a point-like theory of quantum gravity could
have an ultraviolet-finite perturbative expansion.  In particular, we shall 
consider the theory of gravity with the maximal supersymmetry
compatible with having particles of at most spin two ---
the ungauged version of $\NeqEight$
supergravity~\cite{deWitFreedman,CremmerJuliaScherk,CremmerJulia}.

The on-shell ultraviolet divergences of $\NeqEight$ supergravity,
{\it i.e.}~those which cannot be removed by field redefinitions, 
can be probed by studying the ultraviolet behavior of multi-loop
on-shell amplitudes for graviton scattering.  Such scattering amplitudes would
be very difficult to compute in a conventional framework using
Feynman diagrams.  However, tree amplitudes in gravity
can be expressed in terms of tree amplitudes in gauge theory,
by making use of the Kawai-Lewellen-Tye (KLT) relations~\cite{KLT},
or more recent relations found by three of the present authors~\cite{BCJ08}.
Loop amplitudes can be constructed efficiently from tree amplitudes
via generalized
unitarity~\cite{GeneralizedUnitarityOld,ee4partons,%
MultiLoopDDimGenUnitarity,MoreGenUnitarity,BCFUnitarity},
particularly in theories with maximal supersymmetry.
Using these methods, the four-graviton amplitude in $\NeqEight$ supergravity
has been computed at two~\cite{BDDPR},
three~\cite{GravityThree,CompactThree} and (most recently)
four loops~\cite{Neq84,Neq44np}.  Aspects of this program have been
reviewed previously in
refs.~\cite{Bern2002kj,Bern2009kf,Dixon2010gz,Bern2010fy}.

There are many other proposals for making sense of quantum gravity with
point-like particles.  For example, the asymptotic safety
program~\cite{AsymptoticSafety} proposes that the Einstein action
for gravity flows in
the ultraviolet to a nontrivial, Lorentz-invariant fixed point.
It has also been suggested that the ultraviolet theory could break
Lorentz invariance~\cite{Horava}.  In contrast to these two
particular approaches, here we will do conventional perturbation theory
around a (possible) Gaussian fixed point.

The remainder of this article is organized as follows.
In \Sect{CountertermSection} we review what is known about the potential
counterterms for $\NeqEight$ supergravity, based on constraints coming
from both $\NeqEight$ supersymmetry and $E_{7(7)}$ invariance.
In \Sect{ScatAmpSection} we briefly mention the connection between
amplitude divergences in various dimensions and the associated counterterms,
for both $\NeqEight$ supergravity and the (closely related)
$\NeqFour$ super-Yang-Mills theory.  In \Sect{KLTSection} we review
the KLT relations between gravity and gauge tree amplitudes.
In \Sect{UnitaritySection} we review how generalized unitarity permits
the efficient reconstruction of
multi-loop amplitudes from tree amplitudes.  In \Sect{UnitarityKLTSection}
we show how the combination of unitarity and the KLT relations simplifies
the computation of $\NeqEight$ supergravity loop amplitudes, by relating them
to (planar and non-planar) loop amplitudes in $\NeqFour$ super-Yang-Mills theory.
Finally, in \Sect{ResultsSection} we describe the four-graviton amplitudes
that have been determined at two, three and four loops using these methods,
and we discuss their ultraviolet properties.  We present our conclusions
in \Sect{ConclusionSection}.


\section{Counterterm constraints}
\label{CountertermSection}

In field theory, ultraviolet divergences are associated with local
counterterms.  The divergences that survive in on-shell scattering
amplitudes should respect the symmetries of the theory; in theories of
gravity the counterterms should be generally covariant.  Thus they are
expressible as products of the Riemann tensor $R^\mu_{\nu\sigma\rho}$,
along with covariant derivatives ${\cal D}_\mu$ acting on it.  (If
matter is present, then the energy momentum tensor $T_{\mu\nu}$ can also
appear.)  The loop-counting parameter $G_N$ has mass dimension $-2$,
while the Riemann tensor has mass dimension 2: 
$R^\mu_{\nu\sigma\rho}\sim \del_\rho \Gamma^\mu_{\nu\sigma}
\sim g^{\mu\kappa} \del_\rho\del_\nu g_{\kappa\sigma}$.
Therefore, by dimensional analysis, an $L$-loop counterterm has the
generic form (suppressing all Lorentz indices)
${\cal D}^{2(L+1-p)} R^{p}$ for some power $p$.

Nonlinear field redefinitions of the Einstein action allow the removal of
the Ricci tensor $R_{\mu\nu}$ and scalar $R$ from potential counterterms.
After making such redefinitions, the only available one-loop counterterm
(in a theory of pure gravity), $R_{\mu\nu\sigma\rho}R^{\mu\nu\sigma\rho}$, 
is equivalent to the Gauss-Bonnet term.  The latter is a total derivative,
and cannot be generated in perturbation theory.
This fact explains why pure gravity is finite at one
loop, although there are divergences if matter is
present~\cite{tHooftVeltmanGravity}.

In any pure supergravity theory, {\it i.e.}~one in which all states are
related by supersymmetry to the graviton,
there are also no divergences at two loops.  The reason is
that the unique potential counterterm,
$R^3 \equiv R^{\lambda\rho}_{\mu\nu} R^{\mu\nu}_{\sigma\tau}
R^{\sigma\tau}_{\lambda\rho}$,
is incompatible with ${\cal N}=1$ supersymmetry.
When sandwiched between four graviton plane-wave states, $R^3$
produces a nonzero matrix element~\cite{Grisaru,DeserKayStelle,Tomboulis}
for helicity configurations $({\pm}{+}{+}{+})$ that are forbidden by 
supersymmetry Ward identities~\cite{GrisaruSWI}.
Again, if matter super-multiplets are present, other counterterms are available,
and lower-loop divergences are possible, even at one loop~\cite{vanN1976bg}.

In pure supergravity, the first potential counterterm appears at three
loops~\cite{DeserKayStelle,Ferrara1977mv,Deser1978br,%
Howe1980th,Kallosh1980fi}, and is often abbreviated as $R^4$.
It has long been known to be compatible with not just ${\cal N}=1$ supersymmetry,
but the full $\NeqEight$, because it appears as the first subleading term (after
the Einstein action) in the low-energy limit of the four-graviton scattering
amplitude in type II closed superstring theory~\cite{Gross1986iv},
\be
\langle R^4 \rangle |_{4-{\rm point}} =
s t u \, M_4^{\tree}(1,2,3,4) \,.
\label{R4matrixelement}
\ee 
Here the momentum invariants are
$s=(k_1+k_2)^2$, $t=(k_2+k_3)^2$, $u=(k_1+k_3)^2$, and
$M_4^\tree$ stands for any of the $256^4$ four-point amplitudes in
${\cal N}=8$ supergravity (after removing the gravitational coupling constant).
The $R^4$ operator was ruled out as a counterterm for $\NeqEight$ supergravity
by analyzing the ultraviolet behavior of the three-loop four-graviton
amplitude~\cite{GravityThree,CompactThree} (see \Sect{ResultsSection}).

Recently, Elvang, Freedman and Kiermaier~\cite{EFK2010} studied
the constraints of $\NeqEight$ supersymmetry on counterterms of
higher operator dimension, and also with more than four powers 
of the Riemann tensor.
The latter only affect amplitudes with more than four external legs.
The first non-vanishing $n$-graviton tree amplitudes are the
maximally-helicity-violating (MHV) ones, which contain two negative graviton
helicities, and $(n-2)$ positive helicities.
For MHV amplitudes, the supersymmetry Ward
identities~\cite{GrisaruSWI} imply that the amplitudes,
divided by a simple prefactor, are Bose
symmetric~\cite{OneloopMHVGravity}.  All non-vanishing
four-point amplitudes are MHV (for gravitons only, the 
only non-vanishing case is $({-}{-}{+}{+})$).
Therefore $\NeqEight$-supersymmetric on-shell counterterms of the form
${\cal D}^{2k}R^4$ can be classified in terms of Bose-symmetric polynomials
$P_k(s,t,u)$ of degree $k$, where $s+t+u=0$.  This analysis leads to
one independent operator each of the form $R^4$ and ${\cal D}^{2k}R^4$ for
$k=2,3,4,5$, with multiple operators appearing first at order
${\cal D}^{12}R^4$.  By dimensional analysis, ${\cal D}^{2k}R^4$ counterterms
are associated with divergences in $D=4$ at loop order $L=k+3$.
All five-point amplitudes are MHV as well (for gravitons only,
either $({-}{-}{+}{+}{+})$ or its parity conjugate), so the
Bose-symmetry constraints (in more variables) are still valid.
For $n=6,7$, a much more sophisticated analysis of the $\NeqEight$ 
supersymmetry Ward
identities on next-to-MHV amplitudes is required~\cite{Elvang2009wd}.
The upshot is that $\NeqEight$ supersymmetry alone is sufficient to rule
out all counterterms through seven loops except for $R^4$, ${\cal D}^4R^4$
and ${\cal D}^6R^4$.  (Earlier work ruled out the four-loop counterterms
${\cal D}^2R^4$ and $R^5$~\cite{DHHK,Kallosh4loop}.)

However, there is another constraint on counterterms in $\NeqEight$
supergravity in $D=4$, 
and that is invariance under the continuous symmetry $E_{7(7)}$, 
a non-compact form of the exceptional Lie group $E_7$~\cite{CremmerJulia}.
The theory contains 70 massless scalars, which parametrize the coset space
$E_{7(7)}/SU(8)$.  The non-$SU(8)$ part of $E_{7(7)}$ is realized nonlinearly,
through motions on the coset manifold parametrized by the
scalar fields.  Therefore it imposes another set of amplitude Ward identities,
which are associated with soft limits as the momentum of one or more scalar
particles approaches zero~\cite{BEZ,AHCKGravity,Kallosh2008rr}.
For example, in the limit that a single scalar becomes soft,
all the matrix elements of a potential
counterterm should vanish.  If $E_{7(7)}$ is also a symmetry
at the quantum level, then these properties can be used to constrain potential
counterterms.  The $SU(8)$ subgroup of $E_{7(7)}$ was shown to be
non-anomalous at one-loop long ago~\cite{Marcus1985yy}.  More recently,
Bossard, Hillmann and Nicolai~\cite{Bossard2010dq}, using a formulation
for the vector fields that has manifest electric-magnetic duality, 
but is not Lorentz covariant, have extended this result to the full
$E_{7(7)}$ symmetry, and to all orders in perturbation theory.

In ref.~\cite{Brodel2009hu} it was found
that the single-soft-scalar limit was non-vanishing for the operator
$e^{-6\phi}R^4$ generated by string theory, where $\phi$ is the dilaton
(plus terms generated by $\NeqEight$ supersymmetry).
This result suggested that the $R^4$ operator might be ruled out as a
counterterm. A more refined analysis~\cite{Elvang2010kc} isolated the
matrix elements generated solely by the $SU(8)$-singlet operator $R^4$
({\it i.e.}~removing effects of the dilaton), and still found a nonvanishing
single-soft limit, thereby demonstrating at the amplitude
level that the $R^4$ counterterm is not allowed by linearized $E_{7(7)}$.
Later this analysis was extended~\cite{Beisert2010jx} to a large
set of higher-dimension operators, and has served to rule out, via
$E_{7(7)}$, the ${\cal D}^4R^4$ and
${\cal D}^6R^4$ potential counterterms mentioned above, as well
as to constrain the potential seven-loop counterterm to have a unique
form, corresponding to ${\cal D}^8R^4$ (plus terms generated by $\NeqEight$
supersymmetry; see ref.~\cite{Freedman2011uc}).  In other words, the
seven-loop finiteness of $\NeqEight$ supergravity in $D=4$ can be
assessed purely by computing the four-graviton scattering amplitude.
Similar conclusions about the finiteness of $\NeqEight$ supergravity
through seven loops (as well as results concerning first divergences in
higher-dimensional versions of the theory), were arrived at in
ref.~\cite{Bossard2010bd},
based also on the $E_{7(7)}$ invariance of counterterms, but using three
different lines of analysis, including the dimensional reduction of
higher-dimensional counterterms. 

A seven-loop, $\NeqEight$ supersymmetric counterterm was constructed
long ago~\cite{Howe1980th}, but that construction was not manifestly
$E_{7(7)}$ invariant.  More recently it was found~\cite{Bossard2009mn}
that this candidate counterterm can be identified with the volume of
the on-shell $\NeqEight$ superspace, and that it is $E_{7(7)}$ invariant,
although it is still possible that it vanishes after using the
classical field equations.
However, it seems more likely that the volume coincides
with the ${\cal D}^8R^4$ potential counterterm that passes the
$E_{7(7)}$ constraints also studied in ref.~\cite{Beisert2010jx}.  On the
other hand, ref.~\cite{Kallosh2010kk} has discussed the constraints
from $E_{7(7)}$ in the context of a light-cone superspace approach,
and argues that the theory is perturbatively finite to all loop
orders.


\section{Four-graviton scattering amplitudes}
\label{ScatAmpSection}

Even if a counterterm is allowed by all known symmetries, that does not
necessarily
mean that its coefficient is nonzero.  Only an explicit computation can
determine this property for certain.  Seven-loop four-graviton scattering
amplitudes are still a bit beyond present technology.  However, 
the four-loop amplitude can, and has been, computed~\cite{Neq84}, 
and furthermore it also allows access to the ${\cal D}^8R^4$ potential
counterterm, albeit in a different spacetime dimension.

In general, we can test the ultraviolet behavior of the four-graviton
scattering amplitude in $\NeqEight$ supergravity at any loop order
$L$ by increasing the spacetime dimension $D$ associated with
the loop-momentum integration, until the amplitude starts to diverge.
It is instructive to compare this behavior with the corresponding
behavior of the maximally supersymmetric gauge theory, $\NeqFour$
super-Yang-Mills theory ($\NeqFour$ sYM).   The latter theory is
known to be finite to all loop orders in $D=4$~\cite{Neq4Finite}.
However, it diverges in $D>4$.  The critical dimension $D_c(L)$,
in which the theory first diverges as $D$ increases,
depends on the number of loops, and
is given by the formula~\cite{BDDPR},
\be
D_c(L)|_{\NeqFour\,{\rm sYM}}\ =\ 4 + \frac{6}{L}\qquad (L>1).
\label{Neq4DcL}
\ee
The surprising result from the four-graviton computations
to be described below, is that, through four loops, $\NeqEight$
supergravity is just as well behaved,
\be
D_c(L)|_{\NeqEight\,{\rm SUGRA}}\ =\ 4 + \frac{6}{L}\qquad (L=2,3,4).
\label{Neq8DcL}
\ee
In both theories, the one-loop case is special, and the first
divergence is in eight dimensions ($D_c(1)=8$).
Clearly, the equality between~(\ref{Neq4DcL}) and (\ref{Neq8DcL}) must
break at some point, if $\NeqEight$ supergravity is to diverge
in four dimensions.

For $\NeqFour$ sYM, the divergences in the critical dimension are
all associated with a single type of counterterm, for $L>1$,
of the general form ${\cal D}^2 F^4$, where $F$ is the gluon field strength
and the color structure is generic.
Given this fact, and recalling that the loop-counting parameter for
gauge theory is dimensionless, while that for gravity is dimensionful,
$G_N = 1/M_{\rm Pl}^2$, the only way that the two formulas for $D_c(L)$
can coincide, is if each successive $\NeqEight$
supergravity divergence in the critical dimension
for $L=2,3,4$ is associated with a counterterm with two more derivatives
(additional powers of the curvature beyond $R^4$ would not produce
a divergence in the four-graviton amplitude).  Indeed, the associated
higher-dimensional counterterms have the form ${\cal D}^{2L}R^4$, $L=2,3,4$.
Thus the ${\cal D}^8R^4$ potential counterterm would correspond to
the divergence of the four-loop four-graviton scattering amplitude
in $D_c(4)=5.5$.  (We do not yet know for sure whether the
amplitude diverges in this dimension; we do know that it does not
diverge in lower dimensions.)

Furthermore, when the divergence in the five-loop amplitude is computed, 
one of two things must happen:  Either (1) the equality of eqs.~(\ref{Neq4DcL})
and (\ref{Neq8DcL}) must break, or else (2) the appropriate operator for
describing the five-loop divergence in the critical dimension
must be ${\cal D}^{10}R^4$.  Because this operator has two {\it more}
derivatives on it than the potential seven-loop counterterm in $D=4$,
${\cal D}^{8}R^4$, possibility (2) would be a strong indicator
that this counterterm is not present.  On the other hand, there have
been predictions, based on the general structure of contributions 
in a world-line formalism using the non-minimal pure spinor
formalism~\cite{Bjornsson2010wm,Bjornsson2010wu}, that 
the equally good ultraviolet behavior of $\NeqEight$ supergravity
and $\NeqFour$ sYM will break at five loops.  
Clearly the ultraviolet behavior of the five-loop
four-graviton amplitude is an important outstanding question, which
will shed strong light on the potential seven-loop divergence
of $\NeqEight$ supergravity in four dimensions.

In the remainder of this contribution, we will outline the technical
tools that have made possible the computation of the complete
four-graviton scattering amplitude in $\NeqEight$
supergravity through four loops, as well as the
extraction of its ultraviolet divergence in the appropriate
higher spacetime dimensions.


\section{The KLT relations}
\label{KLTSection}

As mentioned in the introduction, tree amplitudes in gravity
can be expressed in terms of tree amplitudes in gauge theory,
specifically as bilinear combinations of gauge amplitudes.
The reason this will prove so useful to us is that, by using
generalized unitarity, we will be able to chop the gravity
loop amplitudes up into products of gravity trees.  Then we can
use the gravity-gauge relations to write everything in terms
of products of gauge-theory trees, products which actually appear
in cuts of gauge loop amplitudes.  In this way, multi-loop gauge
amplitudes provide the information needed to construct multi-loop
gravity amplitudes.

The original gravity-gauge tree amplitude relations were found by 
Kawai, Lewellen and Tye~\cite{KLT}, who recognized that
the world-sheet integrands needed to compute tree-level amplitudes
in the closed type II superstring theory were essentially the square
of the integrands appearing in the open-superstring tree amplitudes.
KLT represented the closed-string world-sheet integrals over the complex
plane as products of contour integrals, and then deformed the contours until
they could be identified as integrals for open-string amplitudes,
thus deriving relations between closed- and open-string tree amplitudes.

Because the low-energy limit of the perturbative sector of the
closed type II superstring in $D=4$
is $\NeqEight$ supergravity, and that of the open superstring is
$\NeqFour$ sYM~\cite{Green1982sw}, 
as the string tension goes to infinity the KLT relations 
express any $\NeqEight$ supergravity tree amplitude in terms
of amplitudes in $\NeqFour$ sYM.
More recently, there have been a variety of studies of ``KLT-type''
relations from various perspectives~\cite{KLTtype}.
One set of relations, found by three of the present
authors~\cite{BCJ08}, follows from~\cite{BDHK10}
a color-kinematic duality satisfied by gauge theory amplitudes.
These relations promise to greatly simplify future computations of 
$\NeqEight$ supergravity loop
amplitudes~\cite{BCJ10,CJtoappear,BCDJRtoappear}.
However, in this article we will only describe the use of the KLT
relations, because those were employed in the two-, three-
and four-loop supergravity computations reviewed here.

The KLT relations for $\NeqEight$ supergravity amplitudes are bilinear
in the $\NeqFour$ sYM amplitudes, for two complementary reasons: 
(1) Integrals over the complex plane naturally 
break up into pairs of contour integrals, 
and (2) the $\NeqEight$ supergravity Fock space naturally factors
into a product of ``left'' and ``right'' $\NeqFour$ sYM Fock spaces,
\be
[\NeqEight]\ =\ [\NeqFour]_L \, \otimes \, [\NeqFour]_R \,.
\label{Neq8from4}
\ee
The $256 = 16^2$ massless states of $\NeqEight$ supergravity are
tabulated in the upper half of \Tab{MultiplicityTable}.  Each state can be
associated with a unique pair of states from $\NeqFour$ sYM, which
has 16 massless states (excluding color degrees of freedom),
tabulated in the lower half of the table.  For example, the eight helicity 
$+3/2$ gravitino states are products of helicity $+1$ gluon and helicity $+1/2$
gluino states in two possible ways:  $\psi_A^+ = g^+ \otimes \lambda_A^+$,
$\psi_{A+4}^+ = \lambda_A^+ \otimes g^+$, $A=1,2,3,4$.

\begin{table}[t]
\begin{center}
\begin{tabular}{|c||c|c|c|c|c|c|c|c|c|}
\hline
\multicolumn{10}{|c|}{$\NeqEight$ supergravity} \\
\hline
$h$ & $-2$ & $-\textstyle{\frac{3}{2}}$ & $-1$ &
$-\textstyle{\frac{1}{2}}$ & $0$ & $\textstyle{\frac{1}{2}}$ 
& $1$ & $\textstyle{\frac{3}{2}}$ & $2$ \\  [2pt]
\hline
\# of states & $1$ & $8$ & $28$ & $56$ & $70$ & $56$ & $28$ & $8$ & $1$
\\
\hline
field & $h^-$ & $\psi_i^-$ & $v_{ij}^-$ & $\chi_{ijk}^-$
& $s_{ijkl}$ & $\chi_{ijk}^+$ & $v_{ij}^+$ & $\psi_i^+$ & $h^+$
\\ [2pt]
\hline
\hline
\multicolumn{10}{|c|}{$\NeqFour$ super-Yang-Mills} \\ 
\hline
$h$ & ~ & ~ & $-1$ &
$-\textstyle{\frac{1}{2}}$ & $0$ & $\textstyle{\frac{1}{2}}$ 
& $1$ & ~ & ~ \\ [2pt]
\hline
\# of states & ~ & ~ & $1$ & $4$ & $6$ & $4$ & $1$ & ~ & ~
\\
\hline
field &  &  & $g^-$ & $\lambda_A^-$ & $\phi_{AB}$
& $\lambda_A^+$ & $g^+$ & & 
\\ [2pt]
\hline
\end{tabular}
\end{center}
\caption{\small
Table of state multiplicities, as a function of helicity $h$,
for the $2^8=256$ states in $\NeqEight$ supergravity and for the 
$2^4 = 16$ states in $\NeqFour$ super-Yang-Mills theory. }
\label{MultiplicityTable} 
\end{table}


In the open string theory, color degrees of freedom for gluons
appear as Chan-Paton factors, but these factors 
are not present in the closed string.  Hence
the gauge theory amplitudes appearing in the KLT relations are those
from which the Chan-Paton factors have been stripped off, which
are known in the QCD community as color-ordered
subamplitudes (see {\it e.g.}~ref.~\cite{LDTASI} for a review).
The full color-dressed gauge-theory tree amplitude ${\cal A}_n^\tree$
is given as a sum over permutations of the color-ordered
subamplitudes $A_n^\tree$,
\be
{\cal A}_n^\tree(\{k_i,a_i\}) = g^{n-2} \,
 \sum_{\rho\in S_n/\ZZ_n}
\Tr( T^{a_{\rho(1)}} T^{a_{\rho(2)}} \ldots T^{a_{\rho(n)}} ) \,
 A_n^\tree(\rho(1), \rho(2), \ldots, \rho(n))\,,
\label{Antreedef}
\ee
where $g$ is the gauge coupling, $a_i$ is an adjoint index,
$T^{a_i}$ is a generator matrix in the fundamental representation of
$SU(N_c)$, the sum is over all $(n-1)!$ inequivalent (non-cyclic)
permutations $\rho$ of $n$ objects, and the argument $i$
of $A_n^\tree$ labels both the momentum $k_i$ and state information
(helicity $h_i$, {\it etc.}).

In the case of supergravity tree amplitudes, ${\cal M}_n^\tree$,
only powers of the gravitational coupling $\kappa$ have to be stripped
off, where $\kappa$ is related to Newton's constant by
$\kappa^2 = 32 \pi^2 G_N$.  We define $M_n^\tree$ by
\be
{\cal M}_n^\tree(\{k_i\})
\ =\ \left(\frac{\kappa}{2}\right)^{n-2}\,M_n^\tree(1,2,\ldots,n) \,.
\label{Mntreedef}
\ee
Then the first few KLT relations have the form,
\bea
M_3^\tree(1,2,3) &=& i \, A_3^\tree(1,2,3) \tilde{A}_3^\tree(1,2,3) \,,
\label{KLTThree} \\
M_4^\tree(1,2,3,4) &=&  
     - i s_{12} \, A_4^\tree(1,2,3,4) \, \tilde{A}_4^\tree(1,2,4,3)\,, 
\label{KLTFour} \\
M_5^\tree(1,2,3,4,5) &=& i s_{12} s_{34} \, A_5^\tree(1,2,3,4,5)
                                     \tilde{A}_5^\tree(2,1,4,3,5)
\ +\ \CP(2,3) \,,
\label{KLTFive} \\
M_6^\tree(1,2,3,4,5,6) &=& - i s_{12}s_{45} 
\, A_6^\tree(1,2,3,4,5,6) \nonumber\\
&&\quad
\times \left[ s_{35} \, \tilde{A}_6^\tree(2,1,5,3,4,6)
            + (s_{34}+s_{35}) \, \tilde{A}_6^\tree(2,1,5,4,3,6) \right]
\nonumber\\
&&\quad\  +\ \CP(2,3,4) \,,
\label{KLTSix}
\eea
where $s_{ij} \equiv (k_i+k_j)^2$, and 
``$+\,\CP$'' indicates a sum over the $m!$ permutations of the 
$m$ arguments of $\CP$.  Here $A_n^\tree$ indicates a tree amplitude
for which the external states are drawn from the left-moving Fock space
$[\NeqFour]_L$ in the tensor product~(\ref{Neq8from4}), while
$\tilde{A}_n^\tree$ denotes an amplitude from the right-moving copy
$[\NeqFour]_R$.


\section{Generalized unitarity}
\label{UnitaritySection}

\begin{figure}[t]
\centerline{\epsfxsize 5.2 truein \epsfbox{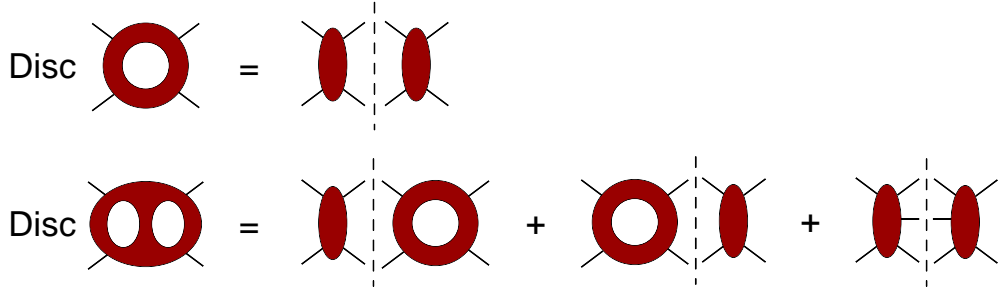}}
\caption[a]{\small Unitarity relations for the four-point amplitude
at one and two loops.  The number of holes in a blob indicates the number
of loops in the corresponding amplitude.}
\label{UnitarityFigure}
\end{figure}

The scattering matrix is a unitary operator between in and out states:
$S^\dagger S = 1$, or in terms of the more standard
``off-forward'' scattering matrix, $T \equiv (S-1)/i$, 
\be
2 \,\hbox{Disc} \, T = T^\dagger T \,,
\label{Tunit}
\ee
where Disc\,$T \equiv (T-T^\dagger)/2i$.
This simple relation generates the well-known unitarity relations,
or cutting rules~\cite{Cutting},
for the discontinuities (or absorptive parts) of
perturbative amplitudes.  If one inserts a perturbative expansion for
$T$ into \eqn{Tunit}, say
\bea
T_4 &=& g^2 \, T_4^\tree + g^4 \, T_4^\oneloop + g^6 \, T_4^\twoloop +
\ldots \,,
\label{T4} \\
T_5 &=& g^3 \, T_5^\tree + g^5 \, T_5^\oneloop + g^7 \, T_5^\twoloop
+ \ldots \,,
\label{T5}
\eea
for the four- and five-point amplitudes, then one obtains the unitarity
relations shown in \Fig{UnitarityFigure}.

At order $g^4$, the discontinuity
in the one-loop four-point amplitude is given by the product of two 
order $g^2$ four-point tree amplitudes.  The product must be summed over 
all possible intermediate states crossing the cut (indicated by 
the dashed line in \Fig{UnitarityFigure}),
and integrated over all possible intermediate momenta.
At two loops, or order $g^6$, there are two possible types of cuts:
the product of a tree-level and a one-loop four-point amplitude
($g^2 \times g^4$), and the product of two tree-level five-point
amplitudes ($g^3 \times g^3$). 

To get the complete scattering amplitude, not just the absorptive part,
one could try to reconstruct the real part via a dispersion relation.
However, in the context of perturbation theory, an easier method is
available, because one knows that the amplitude could have been calculated
in terms of Feynman diagrams.  Therefore it can be expressed as a linear
combination of appropriate Feynman integrals, with coefficients that are
rational functions of the kinematic variables.  The unitarity
method~\cite{UnitarityMethod} matches the information coming from the
cuts against the set of available loop integrals in order to determine
these rational coefficients.  Using unitarity in $D=4-2\e$
dimensions~\cite{DDimUnitarity,MultiLoopDDimGenUnitarity},
one can also determine the so-called ``rational terms'', 
which have no cuts in $D=4$.

\begin{figure}[t]
\centerline{\epsfxsize 5.2 truein \epsfbox{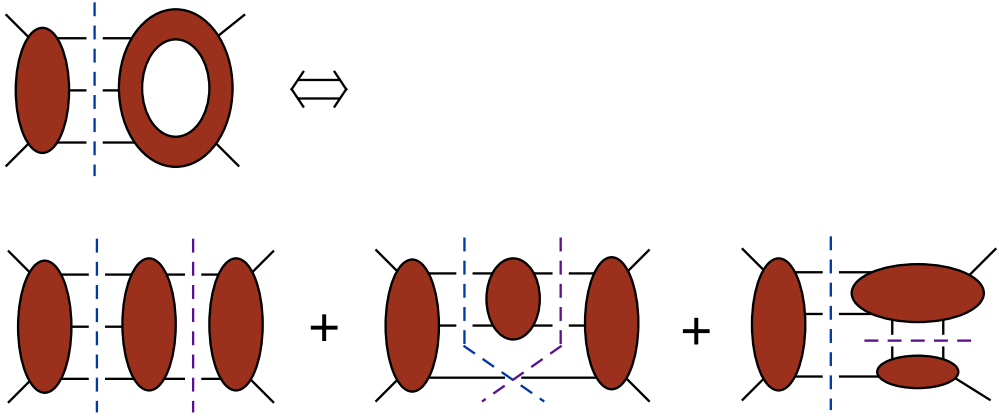}}
\caption[a]{\small An example of multi-loop generalized unitarity.
The one-loop five-point amplitude, appearing on the right side of
the ordinary cut, is further cut into products of trees,
in three inequivalent ways.}
\label{GenUMLFigure}
\end{figure}

{\it Generalized unitarity}~\cite{GeneralizedUnitarityOld}
consists of imposing more than the minimal number of cut lines.
It often simplifies enormously the information required to compute many
terms in the amplitude~\cite{ee4partons,MoreGenUnitarity,BCFUnitarity,%
MultiLoopDDimGenUnitarity},
especially in highly supersymmetric theories~\cite{FourLoop,%
GravityThree,FiveLoop,LeadingSingularity}.
\Fig{GenUMLFigure} provides an example of generalized unitarity at
the multi-loop level.  One starts with an ordinary
three-particle cut for a three-loop four-point amplitude.
The information in this cut can be extracted more easily by cutting
the one-loop five-point amplitude on the right-hand side of the cut, 
decomposing it into the product of a four-point tree and a 
five-point tree, in three inequivalent ways. 

\begin{figure}[t]
\centerline{\epsfxsize 5.2 truein \epsfbox{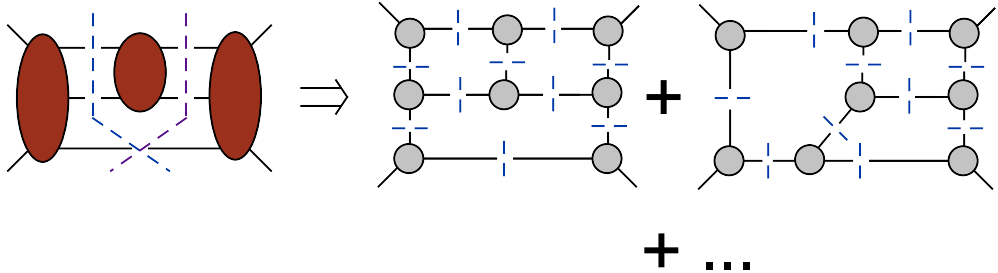}}
\caption[a]{\small A generalized cut with real momenta generates
several maximal cuts; the latter contain only three-point tree amplitudes.}
\label{Cut3ToMaximalFigure}
\end{figure}

\Fig{GenUMLFigure} illustrates a particular class of generalized
unitarity cuts, in which all cut momenta are allowed to be real.  
It is possible, however, to impose more and more on-shell constraints
on intermediate legs, dissolving the amplitude into products of more
tree amplitudes, each with fewer legs (and hence simpler).
For four-point amplitudes,
the maximal cuts are the limiting cases in which all tree amplitudes
are three-point ones, which can be dissolved no further.
\Fig{Cut3ToMaximalFigure} shows how one of the real-momentum
configurations in \Fig{GenUMLFigure} generates several maximal cuts
(which contain complex momenta).
The {\it method of maximal
cuts}~\cite{FiveLoop,CompactThree,Neq44np,CJtoappear}
for constructing a multi-loop amplitude begins with the evaluation
of the maximal cuts, and the construction of a candidate ansatz
for the loop-momentum integrand that is consistent with them.
For simplicity we will discuss here the evaluation of four-dimensional cuts,
that is, cuts in which the cut loop momenta are taken to be in four
dimensions.  For complete generality the cut loop momenta should be
in $D$ dimensions.  However, for the four-point amplitudes in maximally
supersymmetric gauge theory or gravity, the $D$-dimensional cuts have
yet to reveal any new terms, beyond those found using the four-dimensional
cuts~\cite{Neq44np}.

For real momenta, the kinematics of the three-point process
with all massless legs is singular --- all three momenta must be parallel.
However, for complex momenta
it is perfectly nonsingular~\cite{GoroffSagnotti,WittenTwistor}.
The maximal cuts for four-point amplitudes
are enumerated simply by drawing all cubic graphs.
Their evaluation is also very simple, for four-dimensional cuts,
because three-point tree amplitudes are always given by
a simple expression in the usual spinor products, in either 
$\spa{i}.{j} = \pol_{\alpha\beta} \lambda_i^\alpha \lambda_j^\beta$
or $\spb{i}.{j} = \pol_{\dot\alpha\dot\beta}
 \tilde\lambda_i^{\dot\alpha} \tilde\lambda_j^{\dot\beta}$,
where $\lambda_i^\alpha$ ($\tilde\lambda_i^{\dot\alpha}$) is the two-component
positive-chirality (negative-chirality) spinor associated with the massless
momentum $k_i$.  For example, for three gluons there are
only two non-vanishing amplitudes,
\be
A_3^\tree(1^-,2^-,3^+) = i \, { {\spa1.2}^4\over\spa1.2\spa2.3\spa3.1 } \,,
\qquad 
A_3^\tree(1^+,2^+,3^-) = -i \, { {\spb1.2}^4\over\spb1.2\spb2.3\spb3.1 } \,.
\label{mmpppm}
\ee
There are two types of three-point complex kinematics; for each type,
one of the two amplitudes in \eqn{mmpppm} is non-vanishing and the other one
vanishes~\cite{WittenTwistor,BCFUnitarity}.
Three-point amplitudes for gravity can be obtained directly as products
of two gauge amplitudes, using \eqn{KLTThree}.

Even though the maximal cuts are very simple to evaluate analytically,
they provide a great deal of information, and an ansatz that satisfies
the maximal cuts is an excellent starting point for constructing the
full answer.  For example, for the contributions to four-gluon
scattering in $\NeqFour$ sYM that are planar (the dominant terms in the
large $N_c$ limit), the maximal cuts find all terms present in the
amplitude at one, two and three loops.  They only start to miss planar 
terms at four loops (and non-planar terms at three loops).
The remaining terms, whether planar or non-planar,
can be found systematically by collapsing one propagator
in each maximal cut to generate the next-to-maximal cuts; one more propagator
to generate the next-to-next-to-maximal cuts; and so on.  
At each stage the ansatz is improved by adding more terms in order to fit
the new information.  Each additional term should contain at least one
power of an inverse (collapsed) propagator $\ell_i^2$, corresponding to the
fact that it was invisible on the maximal cut ($\ell_i^2=0$),
and only became visible on the next-to-maximal cut ($\ell_i^2\neq0$).
The process of amplitude construction terminates when no more terms
need to be added.  Then the amplitude can be checked, by a comparison
(usually numerical) against a complete, or ``spanning''~\cite{Neq44np},
set of unitarity cuts.


\section{Combining unitarity with KLT}
\label{UnitarityKLTSection}

The general strategy~\cite{BDDPR} we have adopted for computing multi-loop
$\NeqEight$ supergravity amplitudes is to first compute the loop-momentum
integrands for the corresponding amplitudes in $\NeqFour$ sYM.
The integrands are described by a sum of
Feynman integrals for cubic graphs, with standard scalar propagator factors
and additional numerator polynomials. In the four-point case, the $p^{\rm th}$
such integral has the form,
\begin{equation}
{\cal I}^{(p),\,\NeqFour\ {\rm sYM}} =  C^{(p)} \times (-i)^L \int 
\biggl(\prod_{j=1}^L \frac{d^{D} \ell_j}{(2\pi)^D}\biggr) \, 
 \frac{N^{(p)} (\ell_j, k_m)}{\prod_{n=1}^{3L+1} l_n^2} \,,
\label{IntegralNormalization}
\end{equation}
where $k_m$, $m=1,2,3$, are the three independent external momenta,
$\ell_j$ are the $L$ independent loop momenta, and $l_n$ are the
momenta of the $(3L+1)$ propagators (internal lines of the graph $p$),
which are linear combinations of the $\ell_j$ and the $k_m$.
As usual, $d^{D} \ell_j$ is the $D$-dimensional measure for the
$j^{\rm th}$ loop momentum.  The numerator polynomial $N^{(p)}(\ell_j,k_m)$
is a polynomial in both internal and external momenta.
The color factor $C^{(p)}$ can be written as a product of structure constants
$f^{abc}$ for the gauge group.  It can also be written diagrammatically,
using three-vertices for $f^{abc}$ factors, and lines (propagators) for
$\delta^{ab}$ contractions.  In this form, it is given just by the
associated cubic graph.

These integrands can then be cut in any desired
fashion.  Through the KLT relations, they provide the data needed to evaluate
very efficiently the generalized cuts for $\NeqEight$ supergravity.
In particular, the $\NeqEight$ supergravity cuts require a sum over the
256 states in the $\NeqEight$ supergravity multiplet, for every cut line.
However, the corresponding
cut $\NeqFour$ sYM loop integrands already contain a sum
over the 16 states in the $\NeqFour$ sYM multiplet.  The KLT relations
express the $\NeqEight$ supergravity cuts as sums of products of two copies of 
$\NeqFour$ sYM cuts.  The $\NeqEight$ sum factorizes as,
\be 
\sum_{\NeqEight}\ =\ \sum_{[\NeqFour]_L} \sum_{[\NeqFour]_R}\ ,
\ee
and the $\NeqFour$ sums have already been carried out in the course of
constructing the $\NeqFour$ sYM integrand.

Because gravity has no notion of color, planar and non-planar contributions
cannot be separated in graviton amplitudes.
The KLT relations therefore must relate the gravity
cuts to both planar and non-planar gauge theory cuts.  In other words,
the complete $\NeqFour$ sYM amplitude, both planar (large $N_c$) and
non-planar terms, is required in this method.

\begin{figure}[t]
\centerline{\epsfxsize 5.2 truein \epsfbox{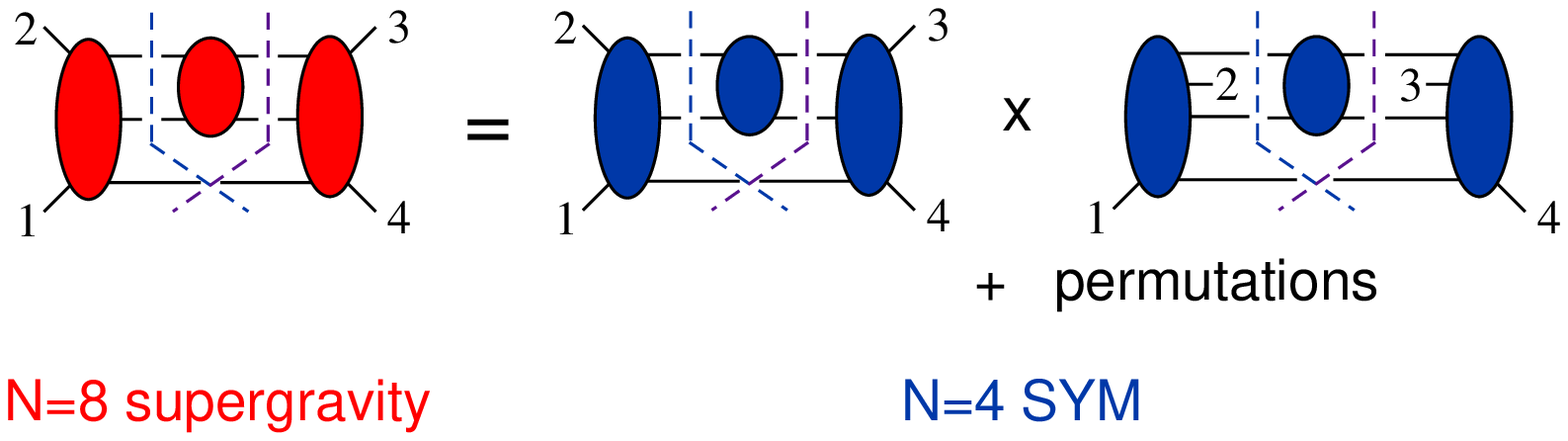}}
\caption[a]{\small Evaluation of a generalized cut in $\NeqEight$
  supergravity at three loops, in terms of planar and non-planar
  cuts in $\NeqFour$ sYM.}
\label{KLTCopyingFigure}
\end{figure}

\Fig{KLTCopyingFigure} sketches how the method works for a particular
generalized cut at three loops.
The $\NeqEight$ supergravity cut contains one four-point tree amplitude
and two five-point ones.  We use the KLT relations~(\ref{KLTFour}) and
(\ref{KLTFive}).  We relabel them, and use the fact that
$s_{12}s_{23} A_4^\tree(1,2,3,4)$ is totally symmetric in legs 1\!,2,3,4
to rewrite them as,
\bea
M_4^\tree(\ell_1,\ell_2,\ell_3,\ell_4) &=&  
     - i {s_{\ell_1\ell_2}s_{\ell_2\ell_3}\over s_{\ell_1\ell_3}}
  \, A_4^\tree(\ell_1,\ell_2,\ell_3,\ell_4)
  \, \tilde{A}_4^\tree(\ell_1,\ell_2,\ell_3,\ell_4)\,, 
\label{KLTNew} \\
M_5^\tree(1,2,\ell_2,\ell_1,\ell_5)
 &=& -i s_{\ell_5 1} s_{2\ell_2} \, A_5^\tree(1,2,\ell_2,\ell_1,\ell_5)
\, \tilde{A}_5^\tree(1,\ell_1,2,\ell_2,\ell_5)
\, + \, \CP(1,2) \,,
\nonumber\\
M_5^\tree(4,3,\ell_3,\ell_4,\ell_5)
 &=& -i s_{\ell_5 4} s_{3\ell_3} \, A_5^\tree(4,3,\ell_3,\ell_4,\ell_5)
 \, \tilde{A}_5^\tree(4,\ell_4,3,\ell_3,\ell_5)
\, + \, \CP(3,4) \,.
\nonumber
\eea
In this way, both occurrences of the four-point $\NeqFour$ sYM
amplitude carry the same cyclic ordering as the $\NeqEight$ supergravity
one, as shown in the figure.  One of the two five-point amplitudes carries
the same ordering, as shown in the left copy.  This copy can be evaluated
using the planar $\NeqFour$ sYM amplitude.  The other five-point amplitude is
twisted, leading to the right copy, which is non-planar, so it requires
non-planar terms in the $\NeqFour$ sYM amplitude.
A reflection symmetry under
the permutation $(1\lr4,\ 2\lr3)$ is preserved by this representation.
The two-fold permutation sum in $M_5^\tree$ in \eqn{KLTNew} leads
to a four-fold permutation sum in the figure; one must add the
permutations $(1\lr2)$, $(3\lr4)$, and $(1\lr2,\ 3\lr4)$.

Note that for terms that are detected
in the maximal cuts, because of the simple relation between
gravity and gauge three-point amplitudes (\eqn{KLTThree}),
the numerator factors are always simply squared in passing from
gauge theory to gravity.


\section{Explicit results}
\label{ResultsSection}

\subsection{Two loops}
\label{TwoLoopSubsection}

\begin{figure}[t]
\centerline{\epsfxsize 4.7 truein \epsfbox{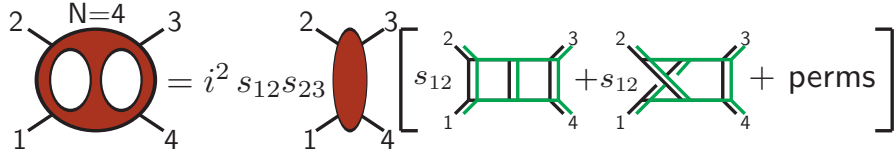}}
\caption[a]{\small The two-loop amplitude in $\NeqFour$ sYM. 
The blob on the right represents the color-ordered tree amplitude 
$A_4^\tree$. In the brackets, black lines are kinematic $1/p^2$ 
propagators, with scalar ($\phi^3$) vertices. 
Green lines are color $\delta^{ab}$ propagators, with structure 
constant ($f^{abc}$) vertices.}
\label{Neq42loopFigure}
\end{figure}
 
The full two-loop four-point amplitude in $\NeqFour$
sYM is given by~\cite{BRY,BDDPR}
\bea
A_4^\twoloop &=& - s_{12} s_{23} A_4^\tree \Bigl[
    C^{\P}_{1234} \, s_{12} \, \I_4^{\twoloop,\P}(s_{12}, s_{23}) 
  + C^{\NP}_{1234} \, s_{12}\, \I_4^{\twoloop,\NP}(s_{12},s_{23}) 
\nonumber\\
&& \hskip2.3cm\null
  + \CP(2,3,4) \Bigr]\,,
\label{TwoLoopYM}
\eea
where $\I_4^{\twoloop,(\P,\NP)}$ are the scalar planar and non-planar
double box integrals shown in \Fig{Neq42loopFigure}, and
$C^{(\P,\NP)}_{1234}$ are color factors constructed from structure 
constant vertices, with the same graphical structure as the corresponding
integral.  The quantity $s_{12} s_{23} A_4^\tree$ is totally symmetric 
under gluon interchange, and its square is the $R^4$ matrix
element in \eqn{R4matrixelement}, up to a factor of $i$.  
Because all terms in \eqn{TwoLoopYM} are detected by the maximal cuts,
the complete two-loop four-point amplitude in $\NeqEight$ supergravity is
found simply by squaring the prefactors in
\eqn{TwoLoopYM} (and removing the color factors, as appropriate for
gravity):
\bea
M_4^\twoloop &=& -i ( s_{12} s_{23} A_4^\tree )^2 \Bigl[
    s_{12}^2 \, \I_4^{\twoloop,\P}(s_{12}, s_{23}) 
  + s_{12}^2 \, \I_4^{\twoloop,\NP}(s_{12},s_{23}) 
  + \CP(2,3,4) \Bigr] \nonumber\\
 &=& s_{12} s_{23} s_{13} M_4^\tree \Bigl[
    s_{12}^2 \, \I_4^{\twoloop,\P}(s_{12}, s_{23}) 
  + s_{12}^2 \, \I_4^{\twoloop,\NP}(s_{12},s_{23}) 
  + \CP(2,3,4) \Bigr]  \,. \nonumber\\
~{} \label{TwoLoopGr}
\eea
Because the loop integrals appearing in the two amplitudes,
\eqns{TwoLoopYM}{TwoLoopGr}, are precisely the same, the critical 
dimension $D_c$ is automatically the same for both theories at two loops.
This value is $D_c=7$, the dimension in which the two-loop,
seven-propagator integrals, $\sim \, \int d^{2D}\ell/(\ell^2)^7$,
are log divergent, in agreement with \eqns{Neq4DcL}{Neq8DcL}.
The two-loop $\NeqEight$ supergravity divergence is associated with
a counterterm of the form ${\cal D}^4R^4$ in $D=7$.
This type of counterterm is permitted by the field-theoretic
duality constraints of ref.~\cite{Bossard2010bd}.

\subsection{Three loops}
\label{ThreeLoopSubsection}

\begin{figure}[t]
\centerline{\epsfxsize 5.5 truein \epsfbox{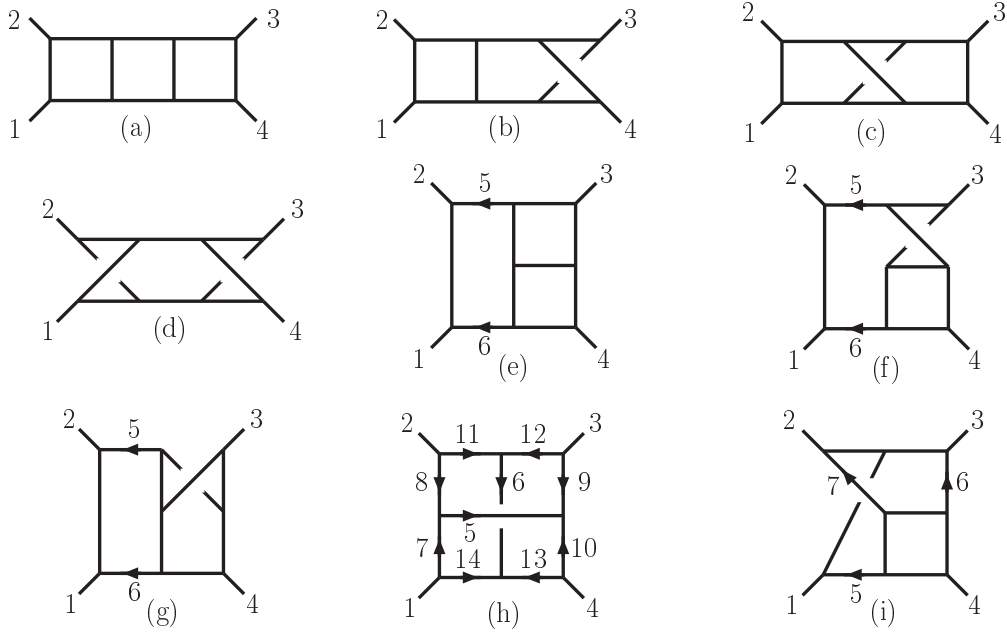}}
\caption[a]{\small Cubic four-point graphs entering the
four-point three-loop amplitudes.}
\label{IntegralsThreeLoopFigure}
\end{figure}

At three loops, the integrand of the $\NeqFour$ sYM four-point amplitude begins to have
dependence on the loop-momentum in its numerator, as well as (non-planar)
terms that cannot be detected in the maximal cuts.  For this reason,
the three-loop $\NeqEight$ supergravity amplitude, in its initial two
forms~\cite{GravityThree,CompactThree}, was not given by simply squaring
the $\NeqFour$ sYM results --- except for a subset of the graphs that could
be inferred using only two-particle cuts.  More recently, three of the present
authors rearranged the three-loop $\NeqFour$ sYM amplitude so as to
make manifest its color-kinematic duality~\cite{BCJ10}.  In this form the
$\NeqEight$ supergravity amplitude can once again be found by a simple 
squaring procedure.
Here we will give the amplitudes in the form found in ref.~\cite{CompactThree},
which requires only the nine cubic graphs shown in
\Fig{IntegralsThreeLoopFigure}.  (Three more cubic graphs, containing
three-point subdiagrams, enter the solution in ref.~\cite{BCJ10}.)

Both the $\NeqFour$ sYM and $\NeqEight$ supergravity amplitudes
are described by giving the loop-momentum numerator polynomials $N^{(p)}$
for these graphs.  In addition, the $\NeqFour$ sYM graphs are multiplied
by the corresponding color structure, as in \Fig{Neq42loopFigure}.

\begin{table}[t]
\begin{center}
\begin{tabular}{||c|c|}
\hline
Integral $I^{(p)}$ & $N^{(p)}$ for $\NeqFour$ super-Yang-Mills \\
\hline
\hline
(a)--(d) &  $
s_{12}^2
$ 
  \\
\hline
(e)--(g) &  $
s_{12} \,s_{46}
$  \\
\hline
(h)&  $\; 
s_{12} (\tau_{26} + \tau_{36}) +
       s_{23} (\tau_{15} + \tau_{25}) + 
       s_{12} s_{23} 
\;$
\\
\hline
(i) & $\;
s_{12} s_{45} - s_{23} s_{46} -  {1\over 3} (s_{12} - s_{23}) \ell_7^2
 \;$  \\
\hline
\end{tabular}
\end{center}
\caption{ \small
The numerator factors $N^{(p)}$ for the integrals $I^{(p)}$ in
\Fig{IntegralsThreeLoopFigure} for $\NeqFour$ super-Yang-Mills theory. 
The first column labels the integral, the second
column the relative numerator factor. An overall factor of 
$s_{12} s_{23} A_4^\tree$ has been removed.  The invariants
$s_{ij}$ and $\tau_{ij}$ are defined in \eqn{InvariantsDef}. }
\label{NumeratorYMTable} 
\vskip .4 cm
\end{table}

\Tab{NumeratorYMTable} gives the values of $N^{(p)}$ for $\NeqFour$ sYM
in terms of the following invariants,
\bea
&& s_{ij} = (k_i +k_j)^2 \,,    \hskip 5.0 cm (i,j \le 4) \nn \\
&& s_{ij} = (k_i +\ell_j)^2 \,,    \hskip 1.5 cm 
\tau_{ij} = 2 k_i\cdot \ell_j\,,   \hskip 1.1 cm (i \le 4,\ \ j\ge5) \nn \\
&& s_{ij} = (\ell_i + \ell_j)^2 \,.    \hskip 5.1 cm  (i, j\ge5) 
\label{InvariantsDef}
\eea
The external momenta $k_i$ are taken to be outgoing in
\Fig{IntegralsThreeLoopFigure}; the directions of the loop momenta $\ell_i$
are indicated by arrows.
Note that $s_{ij}$ is quadratic in the loop momenta $\ell_i$,
if $j>4$, but $\tau_{ij}$ is linear.  Every $N^{(p)}$
in \Tab{NumeratorYMTable} is manifestly quadratic (or better)
in the loop momenta.

\begin{table}[t]
\begin{center}
\begin{tabular}{||c|c||}
\hline
Integral $I^{(p)}$ & $N^{(p)}$ for $\NeqEight$ supergravity  \\
\hline
\hline
(a)--(d) 
& $
\vphantom{\Bigr|}
 [s_{12}^2]^2
$ \\
\hline
(e)--(g) & $\vphantom{\Bigr|}
 s_{12}^2 \, \tau_{35} \, \tau_{46}
$  \\
\hline
(h) & $\;
(s_{12} (\tau_{26} + \tau_{36}) + s_{23}(\tau_{15}+\tau_{25})+s_{12} s_{23})^2
\;$ \\
& $\; \null
   + (s_{12}^2 (\tau_{26} + \tau_{36})
   -  s_{23}^2 (\tau_{15} + \tau_{25})) 
             (\tau_{17} + \tau_{28} + \tau_{39} + \tau_{4,10})
         \; $  \\
    &  $ \; \null
      + s_{12}^2 (\tau_{17} \tau_{28} + \tau_{39} \tau_{4,10})
      + s_{23}^2 (\tau_{28} \tau_{39} + \tau_{17} \tau_{4,10} )
           + s_{13}^2  (\tau_{17} \tau_{39} + \tau_{28} \tau_{4,10}) 
\;     \vphantom{\Bigl( \Bigr)_{A_A} }
         $   \\
\hline
(i) & $
\vphantom{\bigl|_{A_A}} 
(s_{12}\, \tau_{45} - s_{23} \, \tau_{46})^2 
- \tau_{27} (s_{12}^2 \,\tau_{45} + s_{23}^2 \,\tau_{46})
- \tau_{15} (s_{12}^2 \,\tau_{47} + s_{13}^2 \,\tau_{46})
 \; $ \\
& $ \; \null
- \tau_{36} (s_{23}^2\, \tau_{47} + s_{13}^2 \,\tau_{45})
+ \ell_5^2 \, s_{12}^2 \,s_{23} 
+ \ell_6^2 \, s_{12} \, s_{23}^2 
- {1\over 3} \ell_7^2 \, s_{12} \, s_{13} \, s_{23}
\; $ \\
\hline
\end{tabular}
\end{center}
\caption{ \small
Numerator factors $N^{(p)}$ for $\NeqEight$ supergravity.
The first column labels the integral, the second column the relative
numerator factor. An overall factor of 
$s_{12} s_{13} s_{14} M_4^\tree$ has been removed. }
\label{NumeratorGravityTable}
\vskip .4 cm
\end{table}

\Tab{NumeratorGravityTable} gives the values of $N^{(p)}$ for
$\NeqEight$ supergravity, in a form~\cite{CompactThree} which
is also manifestly quadratic in the loop momenta. (In the first version
of the amplitude~\cite{GravityThree}, the quadratic nature was not yet
manifest.)  Comparing the two sets of numerators, we see that the 
$\NeqEight$ supergravity ones are the squares of the $\NeqFour$ sYM
ones, up to contact terms, as expected from the KLT relations.
For example, in graphs (e)--(g), 
$s_{46} = \tau_{46} + \ell_6^2 = \tau_{35} + \ell_5^2$,
so $s_{12}^2 \tau_{35} \tau_{46} \approx [s_{12} s_{46}]^2$ (modulo
$\ell_i^2$ terms).

Because the numerator factors for both $\NeqEight$ supergravity and
$\NeqFour$ sYM are manifestly quadratic in the loop momenta,
the critical dimensions $D_c(L)$ at three loops remain equal,
$D_c(L) = 4+6/L = 6$ for $L=3$.
Indeed, when the ultraviolet poles in the integrals for $\NeqEight$
supergravity are evaluated, no further cancellation is found, and
the resulting pole is
\be
M_4^{\threeloop,\ D=6-2\e} \bigr|_{\rm pole}
\, = \, \frac{1}{\epsilon}\,{5 \, \zeta_3\over (4\pi)^9} \,
(s_{12} s_{23} s_{13})^2 M_4^\tree  \,, \hskip .3 cm 
\label{ThreeLoopD6Div}
\ee
corresponding to a counterterm of the form ${\cal D}^6R^4$ in $D=6$.
Again, the existence of this counterterm is consistent with the field-theoretic
duality constraints of ref.~\cite{Bossard2010bd}.

The form of the divergence~(\ref{ThreeLoopD6Div})
was reproduced from string-theoretic duality
arguments in ref.~\cite{GRV2010}; however, the rational number
predicted there does not agree with \eqn{ThreeLoopD6Div}.
Whether or not this indicates an issue in decoupling massive
states from string theory to obtain $\NeqEight$ 
supergravity~\cite{GOS} remains unclear.

\subsection{Four Loops}
\label{FourLoopSubsection}

\begin{figure}[t]
\centerline{\epsfxsize 5.5 truein \epsfbox{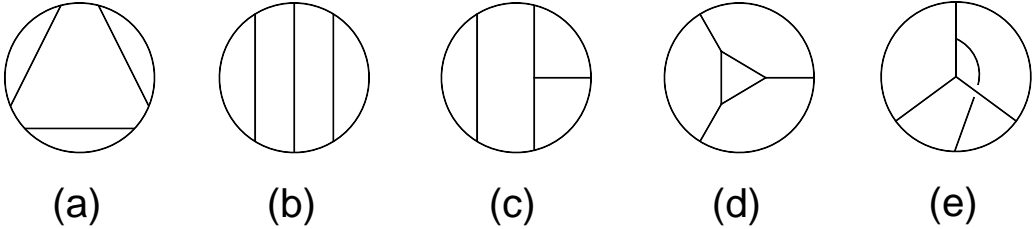}}
\caption[a]{\small Cubic vacuum graphs at four loops.}
\label{vac4Figure}
\end{figure} 

At four loops, the same general strategy still works, but the
bookkeeping issues are greater~\cite{Neq44np}.
One can start by classifying the cubic vacuum graphs.
At three loops there were only two; at four loops there are five,
shown in \Fig{vac4Figure}.  

The next step is to decorate the five vacuum graphs with four external
legs to get the cubic four-point graphs.  As at lower loops, graphs
containing triangles (three propagators or fewer on a loop) or other
three point subgraphs can be dropped.  (This statement would not be
true for representations obeying the color-kinematic duality, as
at three loops~\cite{BCJ10}.)
\Fig{vac4Figure}(a) only gives rise to triangle-containing graphs, so
it can be dropped.
Altogether there are 50 cubic four-point graphs with nonvanishing numerators.
Graphs (b) and (c) do generate four-point graphs without triangles,
but the numerators for all such graphs can be determined, up to possible
contact terms, by iterated two-particle cuts.  Because of the structure
of these cuts~\cite{BRY}, the associated numerator polynomials turn out
to be very simple. Graphs (d), and particularly (e), give rise to the
most complex numerators.

\begin{figure}[tbh]
\centerline{\epsfxsize 5.6 truein \epsfbox{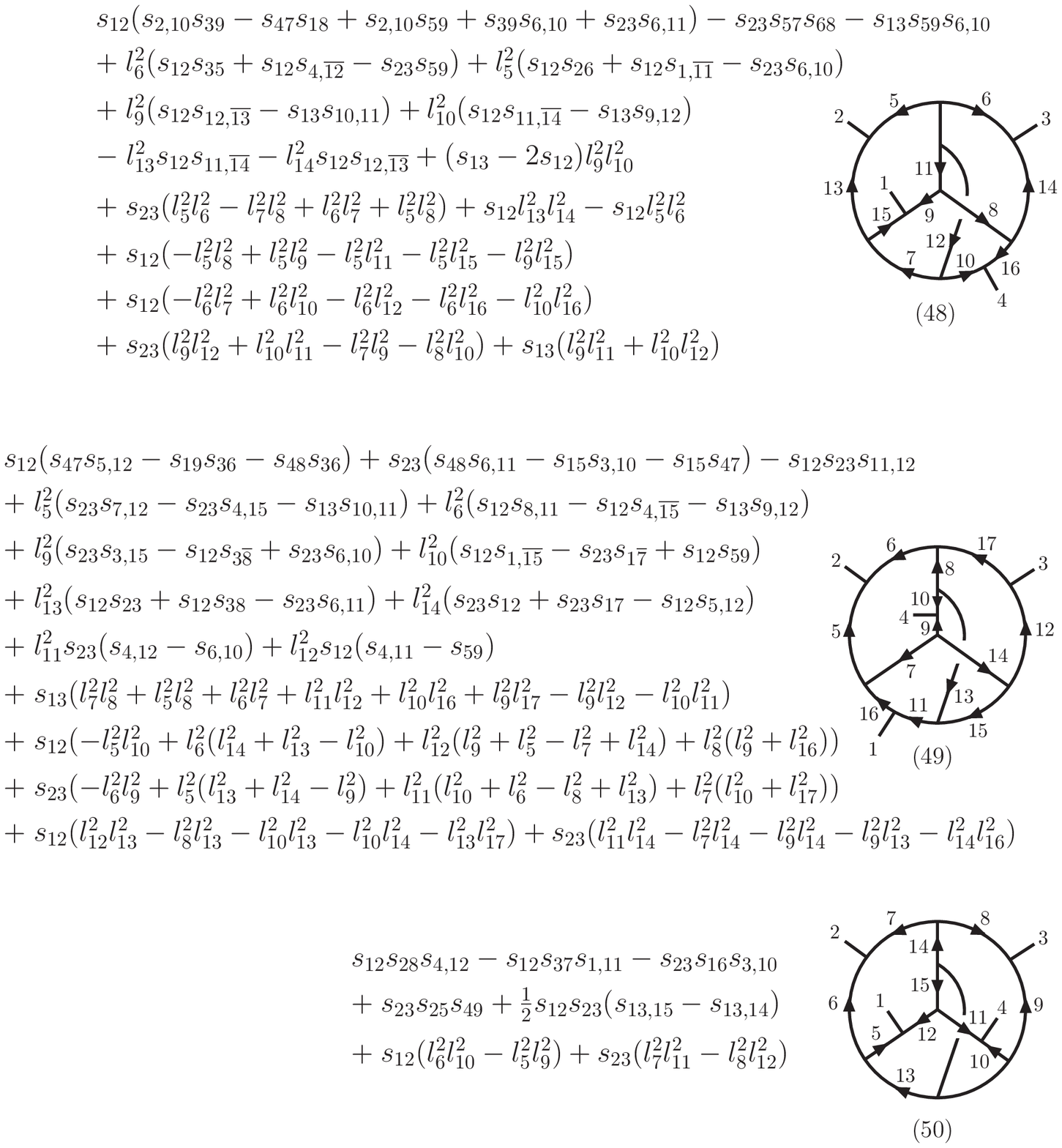}}
\caption[a]{\small Integrals (48)-(50) for the four-loop $\NeqFour$ sYM
  amplitude~\cite{Neq44np}. 
The factor in front of each graph is the numerator polynomial
$N^{(p)}$ for the integral $I^{(p)}$, where $(p)$
is the label below the graph.  The kinematic
variables are defined as $s_{ij}=s_{i,j}=(l_i+l_j)^2$ and
$s_{i,\overline{j}}=(l_i-l_j)^2$,
where $l_i$ is the momentum of line $i$.}
\label{E2Figure}
\end{figure}

The method of maximal cuts was used to determine the numerator polynomials
for $\NeqFour$ sYM.  At four loops, the maximal cuts have 13 cut 
conditions $\ell_i^2=0$.  Then near-maximal cuts with only 12 cut conditions 
are considered, followed by ones with 11 cut conditions.  At this point the 
$\NeqFour$ sYM ansatz is complete; no more terms need to be added.
The result was verified by comparison
against a spanning set of generalized cuts.

In \Fig{E2Figure} we show three of the 50 numerator polynomials.
These three are associated with the one non-planar cubic vacuum graph (e),
and they have the most complex numerators.  Integral (50) is required for the
ansatz for the integrand to match various cuts. However, it integrates
to zero and has vanishing color factor, so it does not contribute to
the $\NeqFour$ sYM amplitude.  In constructing the amplitude, it proved
very useful to have simple pictorial rules that allow one to generate
numerator polynomials for many graphs from those for other graphs, 
either at the same loop order or at lower loop order.
An old rule~\cite{BRY}, called the {\it rung rule}, applies whenever
a graph has a two-particle cut.
A newer rule is the {\it box cut rule}~\cite{FiveLoop,Neq44np}.
It can be applied to any graph that contains a four-point subdiagram,
and it generates that graph's numerator polynomial (modulo certain
contact terms) from the polynomials associated with particular lower-loop
graphs.  Together, these rules are quite powerful; of the 50 graphs, 
only four have neither two-particle cuts nor box cuts.  (Three of the four
appear in \Fig{E2Figure}.)

After the $\NeqFour$ sYM amplitude was computed,
the 50 numerator polynomials for the $\NeqEight$ supergravity amplitude were
then constructed, using information provided by the KLT relations. The results
are quite lengthy, but are provided as {\sc Mathematica} readable files in
ref.~\cite{Neq84}, along with some tools for manipulating them.

From the numerator polynomials for the $\NeqEight$ supergravity amplitude, 
the amplitude's ultraviolet behavior could be extracted,
by expanding the integrals in the limit of small
external momenta, relative to the loop momenta~\cite{TaylorExpand}.
Unlike the three-loop representations in refs.~\cite{CompactThree,BCJ10},
the ultraviolet behavior for the form in ref.~\cite{Neq84} is not manifest.
That means that each integral is more divergent than the sum, and hence
subleading terms in the expansion are required.  It is necessary to expand
to third order, in order to show that $\NeqEight$ supergravity is as well
behaved as $\NeqFour$ sYM at four loops, in this representation
of the amplitude.  More concretely, 
the numerator polynomials, omitting an overall factor of $stu M_4^\tree$,
have a mass dimension of 12, {\it i.e.}~each term is of the form
$k^{12-m} \ell^m$, where $k$ and $\ell$ stand respectively for external
and loop momenta.  The maximum value of $m$ turns out to be 8 for every
integral.  The integrals all have 13 propagators, so they have the form
$\I \sim \int d^{4D}\ell \, \ell^{8-26}$.  The amplitude is manifestly
finite in $D=4$, because $4\times4+8-26<0$.  (This result is not
unexpected, given the absence of a ${\cal D}^2R^4$
counterterm~\cite{DHHK,Kallosh4loop}.)  The amplitude is not
manifestly finite in $D=5$; to see that requires cancellation of the 
$k^4\ell^8$, $k^5\ell^7$ and $k^6\ell^6$ terms, after expansion around
small $k$.

The cancellation of the $k^4\ell^8$ terms is relatively simple, because
one can simply set the external momenta $k_i$ to zero inside
the integrals that appear.  At this point,
the potentially divergent integrals all reduce to one of two types of
scalar vacuum integrals --- there are no loop-momentum tensors appearing
in the numerator, and no doubled propagator factors in the denominator.
In fact, only two of the five vacuum graphs in \Fig{vac4Figure} appear,
(d) and (e).  Collecting all terms, one finds that the coefficients
of (d) and (e) both vanish.  The cancellation
of the $k^5\ell^7$ terms (and the $k^7\ell^5$ terms) is trivial:
Using dimensional regularization, with no dimensionful parameter,
Lorentz invariance does not allow an odd-power divergence.
The most intricate cancellation is that of the $k^6\ell^6$ terms,
corresponding to the vanishing of the coefficient of the potential
counterterm ${\cal D}^6R^4$ in $D=5$.  In the expansion of the integrals
to the second subleading order as $k_i \to 0$, thirty different four-loop 
vacuum integrals are generated.  These integrals often have doubled (and
sometimes tripled) propagators, arising from the Taylor expansion of
the loop-momentum integrand in the external momentum.  Some integrals also
contain tensors in the loop-momentum in their numerators.
However, there are consistency relations
between the integrals, corresponding to the ability to shift the loop
momenta by external momenta before expanding around $k_i = 0$.
These consistency relations are powerful enough to imply the cancellation 
of the ultraviolet pole in $D=5-2\e$.  As a check, we evaluated all 30
ultraviolet poles directly, with the same conclusion.  We did not
yet evaluate the ultraviolet pole near $D=11/2=5.5$ (the critical
dimension for $\NeqFour$ sYM at this loop order), so in principle it
could cancel, although that seems unlikely to be the case.

In summary, the four-loop four-point amplitude of $\NeqEight$ 
supergravity is ultraviolet finite for $D < 11/2$~\cite{Neq84}, the same
bound found for $\NeqFour$ super-Yang-Mills theory.
Finiteness in $5\le D<11/2$ is a consequence of nontrivial cancellations, 
beyond those already found at three loops~\cite{GravityThree,CompactThree}. 
These results provide the strongest direct support to date
for the possibility that $\NeqEight$ supergravity might be a
perturbatively finite quantum theory of gravity.


\section{Conclusions}
\label{ConclusionSection}

In every explicit computation to date, through four loops,
the ultraviolet behavior of $\NeqEight$ supergravity has proven to be
no worse than that of $\NeqFour$ super-Yang-Mills theory.
On the other hand, there are several recent arguments~\cite{Beisert2010jx,%
Bossard2010bd} in favor of the existence of a seven-loop
counterterm~\cite{Howe1980th} of the form ${\cal D}^8R^4$.  As argued in
\Sect{ScatAmpSection}, the five-loop four-graviton scattering amplitude,
when evaluated in higher dimensions for the loop momentum, should provide a
fairly decisive test for what will happen at seven loops.  Although this
computation is difficult, it may well prove feasible using new ideas
related to the color-kinematic
duality~\cite{BCJ08,BCJ10,BDHK10,CJtoappear,BCDJRtoappear}.

Suppose that $\NeqEight$ supergravity turns out to be finite to all orders in
perturbation theory.  This result still would not prove that it is a consistent
theory of quantum gravity at the non-perturbative level.
There are at least two reasons to think that
it might need a non-perturbative ultraviolet completion:
\begin{enumerate}
\item The (likely) $L!$ or worse growth of the coefficients of the 
order $L$ terms in the perturbative
expansion, which for fixed-angle scattering, would imply a non-convergent 
behavior $\sim L! \, (s/M_{\rm Pl}^2)^L$.
\item The fact that the perturbative series seems to be $E_{7(7)}$
invariant, while the mass spectrum of black holes is non-invariant (see
{\it e.g.}~ref.~\cite{BFK} for recent discussions).
\end{enumerate}
QED is an example of a perturbatively well-defined theory that needs 
an ultraviolet completion;
it also has factorial growth in its perturbative coefficients,
$\sim L! \, \alpha^L$, due to ultraviolet renormalons associated with the
Landau pole.  Yet for small values of $\alpha$ QED works extremely
well: it predicts the anomalous magnetic moment of the
electron to 10 digits of accuracy.  Also, there are many pointlike
non-perturbative ultraviolet completions for QED, namely asymptotically free
grand unified theories.  Are there any imaginable pointlike completions
for $\NeqEight$ supergravity?  Maybe the only completion is string theory;
or maybe this cannot happen because of the impossibility of decoupling
non-perturbative string states not present in $\NeqEight$
supergravity~\cite{GOS}.

Another question is whether $\NeqEight$ supergravity might point the
way to other, more realistic finite (or well behaved) theories of
quantum gravity, having less supersymmetry and (perhaps) chiral
fermions.  One step in this direction could be to examine the
multi-loop behavior of theories that can be thought of as
spontaneously broken gauged $\NeqEight$
supergravity~\cite{Ferrara1979fu}, which are known to have improved
ultraviolet behavior at one loop~\cite{Sezgin1981ac}.

In any event, the excellent perturbative ultraviolet behavior of
$\NeqEight$ supergravity has already provided many surprises.
Although the theory may not itself be of direct
phenomenological interest, perhaps it will some day lead
to more realistic theories also having excellent ultraviolet behavior.
As a ``toy model'' for a pointlike theory of quantum gravity, it has been
extremely instructive, and further exploration will no doubt be fruitful
as well.


\section*{Acknowledgments}
\vskip -.3 cm

L.D. thanks the organizers of the XVI$^{\rm th}$ European Workshop on
String Theory in Madrid for the opportunity to present this work,
and G. Dall'Agata, S. Ferrara, and F. Zwirner for useful conversations.
This research was supported by the US Department of Energy
under contracts DE--AC02--76SF00515, DE--FG03--91ER40662 and
DE-FG02-90ER40577 (OJI), by the US National Science Foundation under
grant PHY-0855356, and by the Alfred P. Sloan Foundation.
J.J.M.C. gratefully acknowledges the Stanford Institute for Theoretical
Physics for financial support.
H.J.'s research is supported by the European Research Council
under Advanced Investigator Grant ERC-AdG-228301.
The figures were generated using Jaxodraw~\cite{Jaxo1and2}, based on
Axodraw~\cite{Axo}.



\begin{thebibliography}{99}

\bibitem{deWitFreedman}
B.~de Wit and D.~Z.~Freedman,
Nucl.\ Phys.\  B {\bf 130}, 105 (1977).

\bibitem{CremmerJuliaScherk}
E.~Cremmer, B.~Julia and J.~Scherk,
Phys.\ Lett.\ B {\bf 76}, 409 (1978).

\bibitem{CremmerJulia}
E.~Cremmer and B.~Julia,
Phys.\ Lett.\ B {\bf 80}, 48 (1978);
Nucl.\ Phys.\  B {\bf 159}, 141 (1979).

\bibitem{KLT}
H.~Kawai, D.~C.~Lewellen and S.~H.~H.~Tye,
Nucl.\ Phys.\ B {\bf 269}, 1 (1986).

\bibitem{BCJ08}
Z.~Bern, J.~J.~M.~Carrasco and H.~Johansson,
Phys.\ Rev.\  D {\bf 78}, 085011 (2008)
[0805.3993 [hep-ph]].

\bibitem{GeneralizedUnitarityOld}
R.~J.~Eden, P.~V.~Landshoff, D.~I.~Olive and J.~C.~Polkinghorne,
{\it The Analytic S Matrix} (Cambridge University Press, 1966).

\bibitem{ee4partons}
Z.~Bern, L.~J.~Dixon and D.~A.~Kosower,
Nucl.\ Phys.\ B {\bf 513}, 3 (1998)
[hep-ph/9708239].

\bibitem{MultiLoopDDimGenUnitarity}
Z.~Bern, L.~J.~Dixon and D.~A.~Kosower,
JHEP {\bf 0001}, 027 (2000)
[hep-ph/0001001];\\
%
JHEP {\bf 0408}, 012 (2004)
[hep-ph/0404293].

\bibitem{MoreGenUnitarity}
Z.~Bern, V.~Del Duca, L.~J.~Dixon and D.~A.~Kosower,
Phys.\ Rev.\  D {\bf 71}, 045006 (2005)
[hep-th/0410224].

\bibitem{BCFUnitarity}
R.~Britto, F.~Cachazo and B.~Feng,
Nucl.\ Phys.\  B {\bf 725}, 275 (2005)
[hep-th/0412103].

\bibitem{BDDPR}
Z.~Bern, L.~J.~Dixon, D.~C.~Dunbar, M.~Perelstein and J.~S.~Rozowsky,
Nucl.\ Phys.\ B {\bf 530}, 401 (1998)
[hep-th/9802162].

\bibitem{GravityThree}
Z.~Bern, J.~J.~Carrasco, L.~J.~Dixon, H.~Johansson, D.~A.~Kosower 
and R.~Roiban,
Phys.\ Rev.\ Lett.\  {\bf 98}, 161303 (2007)
[hep-th/0702112].

\bibitem{CompactThree}
Z.~Bern, J.~J.~M.~Carrasco, L.~J.~Dixon, H.~Johansson and R.~Roiban,
Phys.\ Rev.\  D {\bf 78}, 105019 (2008)
[0808.4112 [hep-th]].

\bibitem{Neq84}
Z.~Bern, J.~J.~Carrasco, L.~J.~Dixon, H.~Johansson and R.~Roiban,
Phys.\ Rev.\ Lett.\  {\bf 103}, 081301 (2009)
[0905.2326 [hep-th]].

\bibitem{Neq44np}
Z.~Bern, J.~J.~Carrasco, L.~J.~Dixon, H.~Johansson and R.~Roiban,
Phys.\ Rev.\  {\bf D82}, 125040 (2010)
[1008.3327 [hep-th]].

\bibitem{Bern2002kj}
Z.~Bern,
Living Rev.\ Rel.\  {\bf 5}, 5 (2002)
[gr-qc/0206071].

\bibitem{Bern2009kf}
Z.~Bern, J.~J.~M.~Carrasco and H.~Johansson,
0902.3765 [hep-th].

\bibitem{Dixon2010gz}
L.~J.~Dixon,
1005.2703 [hep-th].

\bibitem{Bern2010fy}
Z.~Bern, J.~J.~M.~Carrasco and H.~Johansson,
Nucl.\ Phys.\ Proc.\ Suppl.\  {\bf 205-206}, 54 (2010)
[1007.4297 [hep-th]].

\bibitem{AsymptoticSafety}
S.~Weinberg, in {\it Understanding the Fundamental Constituents of
Matter}, ed. A.~Zichichi (Plenum Press, New York, 1977);
S.~Weinberg, in {\it General Relativity}, eds. S.~W.~Hawking and W.~Israel
(Cambridge University Press, 1979), p. 700;\\
M.~Niedermaier and M.~Reuter,
Living Rev.\ Rel.\ {\bf 9}, 5 (2006).

\bibitem{Horava}
P.~Ho\v{r}ava,
Phys.\ Rev.\  D {\bf 79}, 084008 (2009)
[0901.3775 [hep-th]].

\bibitem{tHooftVeltmanGravity}
G.~'t Hooft and M.~J.~G.~Veltman,
Annales Poincare Phys.\ Theor.\ A {\bf 20}, 69 (1974).

\bibitem{Grisaru}
M.~T.~Grisaru,
Phys.\ Lett.\  B {\bf 66}, 75 (1977).

\bibitem{DeserKayStelle}
S.~Deser, J.~H.~Kay and K.~S.~Stelle,
Phys.\ Rev.\ Lett.\  {\bf 38}, 527 (1977).

\bibitem{Tomboulis}
E.~Tomboulis,
Phys.\ Lett.\  B {\bf 67}, 417 (1977).

\bibitem{GrisaruSWI}
M.~T.~Grisaru, H.~N.~Pendleton and P.~van Nieuwenhuizen,
Phys.\ Rev.\ D {\bf 15}, 996 (1977);\\
%
M.~T.~Grisaru and H.~N.~Pendleton,
Nucl.\ Phys.\ B {\bf 124}, 81 (1977).

\bibitem{vanN1976bg}
P.~van Nieuwenhuizen and J.~A.~M.~Vermaseren,
Phys.\ Lett.\ B {\bf 65}, 263 (1976).

\bibitem{Ferrara1977mv}
S.~Ferrara and B.~Zumino,
Nucl.\ Phys.\ B {\bf 134}, 301 (1978).

\bibitem{Deser1978br}
S.~Deser and J.~H.~Kay,
Phys.\ Lett.\  B {\bf 76}, 400 (1978).

\bibitem{Howe1980th}
P.~S.~Howe and U.~Lindstr\"om,
Nucl.\ Phys.\  B {\bf 181}, 487 (1981).

\bibitem{Kallosh1980fi}
R.~E.~Kallosh,
Phys.\ Lett.\ B {\bf 99}, 122 (1981).

\bibitem{Gross1986iv}
D.~J.~Gross and E.~Witten,
Nucl.\ Phys.\  B {\bf 277}, 1 (1986).

\bibitem{EFK2010}
H.~Elvang, D.~Z.~Freedman and M.~Kiermaier,
JHEP {\bf 1011}, 016 (2010)
[1003.5018 [hep-th]].

\bibitem{OneloopMHVGravity}
Z.~Bern, L.~J.~Dixon, M.~Perelstein and J.~S.~Rozowsky,
Nucl.\ Phys.\ B {\bf 546}, 423 (1999)
[hep-th/9811140].

\bibitem{Elvang2009wd}
H.~Elvang, D.~Z.~Freedman and M.~Kiermaier,
JHEP {\bf 1010}, 103 (2010)
[0911.3169 [hep-th]].

\bibitem{DHHK}
J.~M.~Drummond, P.~J.~Heslop, P.~S.~Howe and S.~F.~Kerstan,
JHEP {\bf 0308}, 016 (2003)
[hep-th/0305202].

\bibitem{Kallosh4loop}
R.~Kallosh,
JHEP {\bf 0909}, 116 (2009)
[0906.3495 [hep-th]].

\bibitem{BEZ}
M.~Bianchi, H.~Elvang and D.~Z.~Freedman,
JHEP {\bf 0809}, 063 (2008)
[0805.0757 [hep-th]].

\bibitem{AHCKGravity}
N.~Arkani-Hamed, F.~Cachazo and J.~Kaplan,
JHEP {\bf 1009}, 016 (2010)
[0808.1446 [hep-th]].

\bibitem{Kallosh2008rr}
R.~Kallosh and T.~Kugo,
JHEP {\bf 0901}, 072 (2009)
[0811.3414 [hep-th]].

\bibitem{Marcus1985yy}
N.~Marcus,
Phys.\ Lett.\ B {\bf 157}, 383 (1985).


\bibitem{Bossard2010dq}
G.~Bossard, C.~Hillmann and H.~Nicolai,
JHEP {\bf 1012}, 052 (2010)
[1007.5472 [hep-th]]

\bibitem{Brodel2009hu}
J.~Br\"{o}del and L.~J.~Dixon,
JHEP {\bf 1005}, 003 (2010)
[0911.5704 [hep-th]].

\bibitem{Elvang2010kc}
H.~Elvang and M.~Kiermaier,
JHEP {\bf 1010}, 108 (2010)
[1007.4813 [hep-th]].

\bibitem{Beisert2010jx}
N.~Beisert {\it et al.},
Phys.\ Lett.\  B {\bf 694}, 265 (2010)
[1009.1643 [hep-th]].

\bibitem{Freedman2011uc}
D.~Z.~Freedman and E.~Tonni,
1101.1672 [hep-th].

\bibitem{Bossard2010bd}
G.~Bossard, P.~S.~Howe and K.~S.~Stelle,
JHEP {\bf 1101}, 020 (2011)
[1009.0743 [hep-th]].

\bibitem{Bossard2009mn}
G.~Bossard, P.~S.~Howe, K.~S.~Stelle,
Phys.\ Lett.\ B {\bf 682}, 137 (2009)
[0908.3883 [hep-th]].

\bibitem{Kallosh2010kk}
R.~Kallosh,
JHEP {\bf 1012}, 009 (2010)
[1009.1135 [hep-th]].

\bibitem{Neq4Finite}
S.~Mandelstam,
Nucl.\ Phys.\  B {\bf 213}, 149 (1983);\\
%
P.~S.~Howe, K.~S.~Stelle and P.~K.~Townsend,
Nucl.\ Phys.\  B {\bf 214}, 519 (1983);\\
%
L.~Brink, O.~Lindgren and B.~E.~W.~Nilsson,
Phys.\ Lett.\  B {\bf 123}, 323 (1983).

\bibitem{Bjornsson2010wm}
J.~Bj\"{o}rnsson and M.~B.~Green,
JHEP {\bf 1008}, 132 (2010)
[1004.2692 [hep-th]].

\bibitem{Bjornsson2010wu}
J.~Bj\"{o}rnsson,
JHEP {\bf 1101}, 002 (2011)
[1009.5906 [hep-th]].

\bibitem{Green1982sw}
M.~B.~Green, J.~H.~Schwarz and L.~Brink,
Nucl.\ Phys.\ B {\bf 198}, 474 (1982).

\bibitem{KLTtype}
J.~M.~Drummond, M.~Spradlin, A.~Volovich and C.~Wen,
Phys.\ Rev.\  D {\bf 79}, 105018 (2009)
[0901.2363 [hep-th]];\\
N.~E.~J.~Bjerrum-Bohr, P.~H.~Damgaard, B.~Feng and T.~Sondergaard,
Phys.\ Rev.\  D {\bf 82}, 107702 (2010)
[1005.4367 [hep-th];
JHEP {\bf 1009}, 067 (2010)
[1007.3111 [hep-th]];\\
B.~Feng and S.~He,
JHEP {\bf 1009}, 043 (2010)
[1007.0055 [hep-th]];\\
B.~Feng, S.~He, R.~Huang and Y.~Jia,
JHEP {\bf 1010}, 109 (2010)
[1008.1626 [hep-th]];\\
N.~E.~J.~Bjerrum-Bohr, P.~H.~Damgaard, T.~Sondergaard and P.~Vanhove,
JHEP {\bf 1101}, 001 (2011)
[1010.3933 [hep-th]].

\bibitem{BDHK10}
Z.~Bern, T.~Dennen, Y.~t.~Huang and M.~Kiermaier,
Phys.\ Rev.\  D {\bf 82}, 065003 (2010)
[1004.0693 [hep-th]]

\bibitem{BCJ10}
Z.~Bern, J.~J.~M.~Carrasco and H.~Johansson,
Phys.\ Rev.\ Lett.\  {\bf 105}, 061602 (2010)
[1004.0476 [hep-th]].

\bibitem{CJtoappear}
J.~J.~M.~Carrasco and H.~Johansson, to appear.

\bibitem{BCDJRtoappear}
Z.~Bern, J.~J.~M.~Carrasco, L.~J.~Dixon, H.~Johansson and R.~Roiban,
to appear.

\bibitem{LDTASI}
L.~J.~Dixon,
in {\it QCD \& Beyond: Proceedings of TASI '95},
ed. D.\ E.\ Soper (World Scientific, 1996)
[hep-ph/9601359].

\bibitem{Cutting}
L.~D.~Landau,
Nucl.\ Phys.\  {\bf 13}, 181 (1959);\\
S.~Mandelstam,
Phys.\ Rev.\  {\bf 115}, 1741 (1959);\\
R.~E.~Cutkosky,
J.\ Math.\ Phys.\  {\bf 1}, 429 (1960).

\bibitem{UnitarityMethod}
Z.~Bern, L.~J.~Dixon, D.~C.~Dunbar and D.~A.~Kosower,
Nucl.\ Phys.\ B {\bf 425}, 217 (1994)
[hep-ph/9403226];
%
Nucl.\ Phys.\ B {\bf 435}, 59 (1995)
[hep-ph/9409265].

\bibitem{DDimUnitarity}
Z.~Bern and A.~G.~Morgan,
Nucl.\ Phys.\ B {\bf 467}, 479 (1996)
[hep-ph/9511336];\\
%
Z.~Bern, L.~J.~Dixon, D.~C.~Dunbar and D.~A.~Kosower,
Phys.\ Lett.\  B {\bf 394}, 105 (1997)
[hep-th/9611127].

\bibitem{FourLoop}
Z.~Bern, M.~Czakon, L.~J.~Dixon, D.~A.~Kosower and V.~A.~Smirnov,
Phys.\ Rev.\  D {\bf 75}, 085010 (2007)
[hep-th/0610248].

\bibitem{FiveLoop}
Z.~Bern, J.~J.~M.~Carrasco, H.~Johansson and D.~A.~Kosower,
Phys.\ Rev.\  D {\bf 76}, 125020 (2007)
[0705.1864 [hep-th]].

\bibitem{LeadingSingularity}
E.~I.~Buchbinder and F.~Cachazo,
JHEP {\bf 0511}, 036 (2005)
[hep-th/0506126];\\
%
F.~Cachazo and D.~Skinner,
0801.4574 [hep-th];\\
%
F.~Cachazo,
0803.1988 [hep-th];\\
%
F.~Cachazo, M.~Spradlin and A.~Volovich,
Phys.\ Rev.\  D {\bf 78}, 105022 (2008)
[0805.4832 [hep-th]];\\
%
M.~Spradlin, A.~Volovich and C.~Wen,
Phys.\ Rev.\  D {\bf 78}, 085025 (2008)
[0808.1054 [hep-th]].

\bibitem{GoroffSagnotti}
M.~H.~Goroff and A.~Sagnotti,
Phys.\ Lett.\  B {\bf 160}, 81 (1985);
Nucl.\ Phys.\  B {\bf 266}, 709 (1986).

\bibitem{WittenTwistor}
E.~Witten,
Commun.\ Math.\ Phys.\  {\bf 252}, 189 (2004)
[hep-th/0312171].

\bibitem{BRY}
Z.~Bern, J.~S.~Rozowsky and B.~Yan,
Phys.\ Lett.\  B {\bf 401}, 273 (1997)
[hep-ph/9702424].

\bibitem{GRV2010}
M.~B.~Green, J.~G.~Russo and P.~Vanhove,
JHEP {\bf 1006}, 075 (2010)
[1002.3805 [hep-th]].

\bibitem{GOS}
M.~B. Green, H. Ooguri and J.~H. Schwarz,
Phys.\ Rev.\ Lett.\  {\bf 99}, 041601 (2007)
[0704.0777 [hep-th]].

\bibitem{TaylorExpand}
A.~A.~Vladimirov,
Theor.\ Math.\ Phys.\  {\bf 43}, 417 (1980)
[Teor.\ Mat.\ Fiz.\  {\bf 43}, 210 (1980)];\\
%
N.~Marcus and A.~Sagnotti,
Nuovo Cim.\  A {\bf 87}, 1 (1985).

\bibitem{BFK}
M.~Bianchi, S.~Ferrara and R.~Kallosh,
Phys.\ Lett.\ B {\bf 690}, 328 (2010)
[0910.3674 [hep-th]];
%
JHEP {\bf 1003}, 081 (2010)
[0912.0057 [hep-th]].

\bibitem{Ferrara1979fu}
S.~Ferrara and B.~Zumino,
Phys.\ Lett.\ B {\bf 86}, 279 (1979).

\bibitem{Sezgin1981ac}
E.~Sezgin and P.~van Nieuwenhuizen,
Nucl.\ Phys.\ B {\bf 195}, 325 (1982).  

\bibitem{Jaxo1and2}
D.~Binosi and L.~Theussl,
Comput.\ Phys.\ Commun.\ {\bf 161}, 76 (2004)
[hep-ph/0309015];\\
D.~Binosi, J.~Collins, C.~Kaufhold and L.~Theussl,
Comput.\ Phys.\ Commun.\  {\bf 180}, 1709 (2009)
[0811.4113 [hep-ph]].

\bibitem{Axo}
J.~A.~M.~Vermaseren,
Comput.\ Phys.\ Commun.\ {\bf 83}, 45 (1994).

\end{thebibliography}
\end{document}